\documentclass[twocolumn,superscriptaddress,aps,preprintnumbers,amsmath,amssymb,prl,nofootinbib,nobibnotes]{revtex4}

\usepackage{dcolumn}
\usepackage{graphicx}
\usepackage{epstopdf}
\usepackage{bm}
\usepackage{hyperref}
\usepackage{color}
\usepackage{comment}

\begin{document}

\title{
Inverted Oscillators for Testing Gravity-induced Quantum Entanglement
}

\author{Tomohiro~Fujita
}
\email[Author to whom any correspondence should be addressed. Email:~]{fujita.tomohiro@ocha.ac.jp}
\affiliation{Department of Physics, Ochanomizu University, Bunkyo, Tokyo 112-8610, Japan}
\affiliation{Kavli Institute for the Physics and Mathematics of the Universe (Kavli IPMU),WPI, UTIAS, The University of Tokyo, Kashiwa, Chiba 277-8568, Japan}

\author{Youka~Kaku}
\affiliation{Department of Physics, Kobe University, Kobe 657-8501, Japan}

\author{Akira~Matsumura}
\affiliation{Department of Physics, Kyushu University, Fukuoka, 819-0395, Japan}

\author{Yuta~Michimura}
\affiliation{Kavli Institute for the Physics and Mathematics of the Universe (Kavli IPMU),WPI, UTIAS, The University of Tokyo, Kashiwa, Chiba 277-8568, Japan}
\affiliation{Research Center for the Early Universe, The University of Tokyo, Bunkyo, Tokyo 113-0033, Japan}

\begin{abstract}
In the quest for quantum gravity, we have lacked experimental verification, hampered by the weakness of gravity and decoherence. Recently, various experiments have been proposed to verify quantum entanglement induced by Newtonian gravitational interactions. However, they are not yet certainly feasible with existing techniques. To search for a new setup, we compute the logarithmic negativity of two oscillators with arbitrary quadratic potential coupled by gravity. We find that unstable inverted oscillators generate gravity-induced entanglement most efficiently and are most resistant to decoherence from environmental fluctuations. 
As an experimental realization, we propose a setup of the optical levitation of mirrors with the anti-spring effect. To avoid decoherence due to photon shot noise, a sandwich configuration that geometrically creates the anti-spring is promising.
\end{abstract}
\maketitle

\section{I. Introduction}

Quantum gravity has been one of the biggest challenges in modern physics for a long time~\cite{Kiefer2006, Woodard2009}. 
Although standard perturbative quantum gravity is widely believed as the low-energy effective theory, there is no direct evidence to support this hypothesis. It has not even been confirmed whether gravitational fields are in a quantum superposed state. Furthermore, several theories have proposed that gravity is not quantized~\cite{Kafri2014,Diosi1987,Penrose1996,Bahrami2014,Oppenheim2023}. 
Given these circumstances, it is an essential step to examine the quantum nature of gravity in the experimentally accessible low-energy regime, where Newtonian gravity remains a valid effective description~\cite{Bose2022}.
Recently, some experimental proposals were made and have attracted much attention~\cite{Bose2017, Marletto2017}. These experiments aim to measure the quantum entanglement produced by the Newtonian gravitational interaction between two masses. Following them, many proposals using matter-wave interferometers~\cite{Nguyen2020, Chevalier2020, Van2020, Torovs2021, Miki2021,Tilly2021}, mechanical oscillator model \cite{Qvarfort2020, Krisnanda2020}, optomechanical systems~\cite{Balushi2018, Miao2020, Wan2017, Matsumura2020,Datta2021, Miki2022,Kaku2023} 
and their hybrid model~\cite{Carney2021, Carney2022, Streltsov2022, Pedernales2022, Matsumura2022} were studied. 

Among these proposals, the authors of Ref.~\cite{Krisnanda2020} found that entanglement generation occurs simply by trapping two masses in a potential and then releasing them. This is because as their wavefunctions spread out after the release, the near side feels relatively strong gravity and the far side feels weak gravity, resulting in non-local quantum correlations. However, two problems were pointed out in a follow-up paper~\cite{Rijavec2021}: one, as with other proposals, requires strong suppression of environmental noise such as air molecule scattering 
to avoid decoherence until measurable entanglement is produced. The other is that unless the experiment is conducted in space with microgravity, the masses will free-fall after potential release. For three seconds, they will fall more than 40 meters down, making the measurement of the generated entanglement difficult. This free-fall problem is a serious issue because if mechanical support is introduced to prevent the masses from falling, large thermal noise through the support would lead to rapid decoherence.

Furthermore, in common with classical experiments, the weakness of gravity is an obstacle. The dimensionless parameter of gravity acting on two identical oscillators with mass $m$ and angular frequency $\omega$ placed at distance $d$ is
\begin{equation}
    \eta\equiv \frac{2Gm}{\omega^2 d^3} =2.7\times 10^{-13}\,
    \omega_{\rm kHz}^{-2}\left(\frac{m/d^3}{\rm 2~g/cm^3}\right)\,,
    \label{eta estimate}
\end{equation}
where we introduced 
$\omega_{\rm kHz}\equiv \omega/1$kHz and  $m/d^3$ is roughly capped by the density of the oscillators.
Since we will consider an optomechanical setup, we have chosen 2g/${\rm cm}^3$ as our fiducial value, referring to the typical density of the mirror material such as fused silica (2.2 g/cm$^3$) or silicon (2.3 g/cm$^3$).
To obtain a measurable signal, we have to overcome the smallness of $\eta$.

In this letter, we seek a new experimental setup addressing the above problems. 
We first provide a generic theoretical framework to consider the gravitational interaction between oscillators. We consider a pair of oscillators with arbitrary quadratic potentials and investigate when the generated entanglement is maximized. By computing logarithmic negativity, we find that unstable inverted oscillators generate entanglement most efficiently. Their entanglement increases exponentially, which is advantageous for overcoming weak gravity and decoherence.
We then explore the experimental realization of such inverted oscillators. The milligram mass range is known to be promising for experimental verification of the quantum nature of gravity~\cite{Schmole2016, Matsumoto2019,Westphal2021}. In that region, the anti-spring effects in cavity optomechanics are well known and practiced~\cite{Chen2013,Aspelmeyer2014}. To avoid decoherence due to photon shot noise, we will consider a sandwich configuration of mirrors~\cite{Michimura2017}.

\section{II. Oscillators with arbitrary quadratic potential}

In this section, we solve the dynamics of two oscillators with arbitrary quadratic potential and calculate the quantum entanglement between them induced by gravity.
We consider a general Hamiltonian for two oscillators coupled by Newtonian gravity,
\begin{align}
H=\frac{p_1^2}{2m} + \frac{1}{2} k_1 x_1^2
+\frac{p_2^2}{2m} + \frac{1}{2} k_2 x_2^2
-\frac{Gm^2}{d^3}(x_1-x_2)^2,
\end{align}
where $k_i\ (i=1,2)$ is spring constant and we keep only relevant interaction terms
by assuming $d\gg |x_i|$.
Introducing dimensionless variables, $P_i\equiv p_i/\sqrt{\hbar m\omega}$, $X_i\equiv \sqrt{m\omega/\hbar}x_i$, 
the above Hamiltonian is rewritten as
\begin{align}
H=\frac{\hbar\omega}{2}\Big[P_1^2 + \lambda_1 X_1^2
+P_2^2 + \lambda_2 X_2^2
-\eta(X_1-X_2)^2\Big],
\label{Hamiltonian}
\end{align}
where the new constant parameters $\lambda_i\equiv k_i/(m\omega^2)$ specify the potentials. When $\lambda_i$ is $+1, 0,$ and $-1$, the potential becomes harmonic, free and inverted, respectively. Here $\omega$ appeared independent of the other parameters, the meaning of which will be described soon below.

The Heisenberg-Langevin equations for the coupled oscillators read
\begin{equation}
    \dot{X}_i = \omega P_i, 
    \quad
    \dot{P}_i = -\lambda_i\,
    \omega X_i + \omega\eta (X_i-X_j)+\xi_i,  
    \label{EoM ij}
\end{equation}
where $i,j=1$ or $2, $  $i\neq j$, and a random force noise $\xi_i$ is added to incorporate environmental fluctuations, while the small dissipation term is ignored ~\cite{Krisnanda2020,Rijavec2021}. We assume $\xi_i$ is a white noise and its amplitude is characterized by a parameter $\mu$ as
\begin{equation}
    \frac{1}{2}\big\langle \xi_i (t) \xi_j(t')+\xi_i (t') \xi_j(t)\big\rangle
    = \mu \omega \delta(t-t')\delta_{ij}\,.
    \label{mu parameter}
\end{equation}
$\mu$ represents the sum of the various decoherence sources. For instance, it encodes air molecular scattering $\mu_{\rm air}\propto p R^2\sqrt{T}$ and thermal photon interaction $\mu_{\rm ph} \propto R^6T^9$, where $R$ is the radius of the oscillator, $p$ and $T$ are the pressure and temperature of the environment (their full expressions can be found in 
Appendix~B).

Since we consider the quadratic potentials, Eq.~\eqref{EoM ij} is exactly solvable.
The derivation is described in 
Appendix~A.
We introduce a vector of the dynamical variables and a covariance matrix,
\begin{align}
    u_i(t)&=\Big(X_1(t), P_1(t), X_2(t), P_2(t)\Big),
    \label{variable u}
    \\
    \sigma_{ij}(t)&=\frac{1}{2}\langle
    u_i(t)u_j(t)+u_j(t)u_i(t)\rangle\,.
    \label{cov mat}
\end{align}

As an initial condition, we consider $\langle u(0)\rangle=0$ and $\sigma_{ij}(0)=\delta_{ij}/2$. 
To prepare the initial state, we assume that both oscillators are first placed in the ground state of a harmonic potential with a common frequency $\omega$.
After this initial state preparation is completed, the spring constants of the two oscillators are quickly switched to $\lambda_i m \omega^2$.
We take the moment of this switching to define the origin of time, $t = 0$, from which the subsequent dynamics are analyzed.
This initialization can be implemented, by applying an auxiliary trapping potential effective only for $t<0$ to allow the system to relax into its ground state.
This initial state preparation is essential, especially for $\lambda_i \le 0$, where the state is not stable.

Having the solution $\sigma_{ij}(t)$, one can define $\tilde{\sigma}_{ij} (t)$ by flipping the sign of the oscillator's momentum $P_2$ and compute the so-called minimum symplectic eigenvalue $\tilde{\nu}_\text{min}$ from $\tilde{\sigma}_{ij}(t)$. 
The explicit form is
\begin{equation}
\tilde{\nu}_{\rm min} \equiv \left[\frac{1}{2}\left(\tilde{\Sigma}-\sqrt{\tilde{\Sigma}^2-4\det \sigma}\right)\right]^{1/2},
\label{nu}
\end{equation}
where 
$\tilde{\Sigma} \equiv \det \sigma_1+\det \sigma_2 -2 \det \sigma_3$ with the $2\times 2$ matrices  
$\sigma_1$,$\sigma_2$ and 
$\sigma_3$ appearing in the block form of 
$\sigma(t)$,
\begin{equation}
\sigma(t)
=
\begin{bmatrix}
\sigma_1 & \sigma_3 \\
\sigma^\text{T}_3 & \sigma_2 \\
\end{bmatrix}.
\label{block}
\end{equation}
It is known that $\tilde{\nu}_{\rm min}<1/2$ is a necessary and sufficient condition for the two oscillators to be entangled~\cite{Vidal2002, Horodecki1996, Peres1996, Simon2000, Giedke2001}. To quantify the generated entanglement, the logarithmic negativity,
\begin{equation}
E_N \equiv \max\left[0,-\log_2 \left(2\tilde{\nu}_{\rm min}\right)\right]\,,
\label{En def}
\end{equation}
is useful. 
When $E_N>0$, the oscillators are entangled, and larger $E_N$ indicates larger entanglement. 
Entanglement between a mechanical oscillator and microwave has been measured experimentally
with an accuracy of $\mathcal{O}(10^{-2})$~\cite{Palomaki2013}.
Hence, in this letter, we set our target negativity as $E_N=10^{-2}$.

\begin{figure}
\begin{center}
\includegraphics[width=\linewidth]{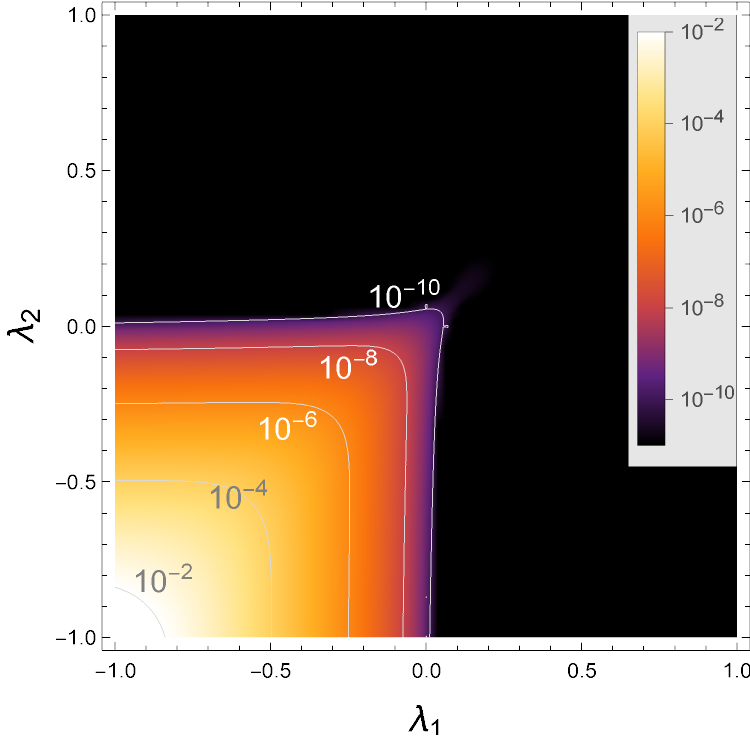}
\caption{The logarithmic negativity $E_N$ of the two oscillator systems induced by the gravitational interaction. We choose $\omega t=13$ and $\eta = 2\mu = 10^{-12}$. 
The horizontal ($\lambda_1$) and vertical ($\lambda_2$) axes denote the potential form of the two oscillators (see Eq.~\eqref{Hamiltonian}) and $\lambda_i=+1, 0,-1$ corresponds to harmonic, free and inverted quadratic potential, respectively.
The largest negativity is generated when both oscillators have the inverted potential (the left bottom corner).}
\label{fig:contour}
\end{center}
\end{figure}
In Fig.~\ref{fig:contour}, we present the logarithmic negativity $E_N$ of the two generic oscillators. 
We observe that a pair of the inverted oscillators generate the largest negativity, while virtually no negativity is produced if either one of the oscillators has positive $\lambda_i$. Since no special point is found in the off-diagonal region of the contour plot, we restrict ourselves to the cases with two identical oscillators by setting $\lambda\equiv \lambda_1=\lambda_2$ (on the diagonal line in Fig.~\ref{fig:contour}) in the rest of this letter.

For two identical oscillators, by expanding the logarithmic negativity with respect to $\eta\sim \mu\ll1$, we obtain a readable result,
\begin{equation}
    E_N(t) \simeq 
    3 \big[\,\eta f_{\rm gra}(t)-\mu f_{\rm dec}(t)\,\big]\,,
    \label{Enandf}
\end{equation}
where the right-hand side is assumed to be positive.
The gravitational term $\eta f_{\rm gra}$ tries to generate the entanglement but the decoherence term $\mu f_{\rm dec}$ tries to prevent it. 
The full expressions for $f_{\rm gra}$ and $f_{\rm dec}$ are given 
in Appendix A.
In the short time limit $\omega t\ll 1$, they marge to $f_{\rm gra}\simeq f_{\rm dec}\simeq \omega t/2$ irrespective of $\lambda$. 
In the long time limit $\omega t\gg 1$, for $\lambda=\pm1$ and 0, we find
\begin{align}
    f_{\rm gra}\simeq \left\{
\begin{array}{ll}
\frac{1}{2}|\sin(\omega t)|\\[3pt]
\frac{1}{6}(\omega t)^3\\[3pt]
\frac{1}{8}e^{2\omega t}
\end{array}
\right.,\ 
f_{\rm dec}\simeq \left\{
\begin{array}{ll}
\frac{1}{2}\omega t& (\lambda =1)\\[3pt]
\frac{1}{6}(\omega t)^3\qquad& (\lambda= 0)\\[3pt]
\frac{1}{8}e^{2\omega t}& (\lambda=-1)
\end{array}
\right. .
\label{fgra eq}
\end{align}
For $\lambda\le 0$, $f_{\rm gra}$ and $f_{\rm dec}$ have the same asymptotic behavior again, and Eq.~\eqref{Enandf} is further approximated by $E_N\simeq 3(\eta-\mu)f_{\rm gra}$. Thus, for entanglement generation, it is necessary to suppress the decoherence effect $\mu$. This fact was known in previous study only for $\lambda=0$ in the short time limit~\cite{Rijavec2021}, but we generalize it for $\lambda\le 0$ in the long time limit.

\begin{figure}
\begin{center}
\includegraphics[width=\linewidth]{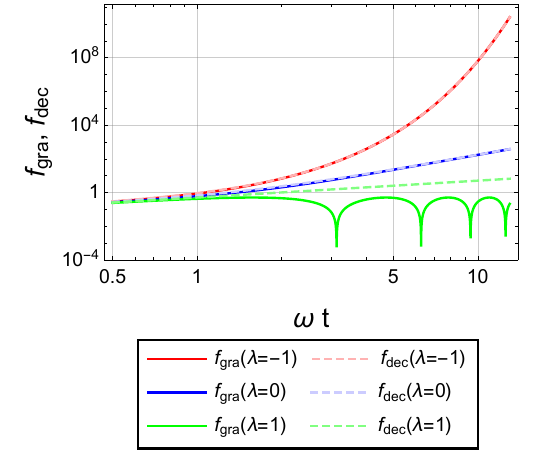}
\caption{
Time evolution of $f_{\rm gra}$ introduced in Eq.~\eqref{Enandf} for $\lambda=-1$ (red), 0 (blue), and 1 (green) against $\omega t$. $f_{\rm dec}$ is also shown as dashed line but almost overlapped with $f_{\rm gra}$ except for $\lambda=1$. $f_{\rm gra}$ converge to $\omega t/2$ in the short time limit irrespective of $\lambda$. In the long time regime, however, the increasing rate of $f_{\rm gra}$ is largely dependent on $\lambda$. The inverted oscillators are particularly prominent for their exponential growth. 
}
\label{fig:fgra}
\end{center}
\end{figure}
Fig.~\ref{fig:fgra} shows the time evolution of $f_{\rm gra}$ and $f_{\rm dec}$ for $\lambda=\pm 1$ and 0. The exponential growth of $f_{\rm gra}$ in the case of the inverted oscillator is remarkable.
To overcome the weakness of gravity ($\eta\ll 1$) and achieve detectable negativity $E_N = 10^{-2}$,  $f_{\rm gra}$ has to quickly increase to a huge value $\simeq 10^{-2}/3\eta$, before the environmental fluctuations decohere the system. 
Combining Eqs.~\eqref{eta estimate}, \eqref{Enandf}, \eqref{fgra eq} and neglecting $\mu$ for a moment, we obtain the time required to generate the detectable entanglement as
\begin{equation}
\tau_{\rm ent} \simeq \left\{
\begin{array}{ll}
4.2\,\omega_{\rm kHz}^{-1/3}\, {\rm sec}\qquad& (\lambda= 0)\\[3pt]
1.3\times 10^{-2}\, \omega_{\rm kHz}^{-1}\, {\rm sec}& (\lambda=-1)
\end{array}
\right..
\label{tau ent}
\end{equation}
The inverted oscillators are roughly three hundred times faster than the free masses at $\omega = 1$kHz and the difference is even greater for larger $\omega$.
Moreover, taking into account the decoherence parameter $\mu$, $\tau_{\rm ent}$ elongates 
by a factor of $[\eta/(\eta-\mu)]^{1/3}$ for $\lambda=0$, while only $\log[\eta/(\eta-\mu)]/(2\omega)$
is added to Eq.~\eqref{tau ent} in the $\lambda=-1$ case. 
For example, when $\mu = \eta/2$ and $\omega = 1$kHz, these correspond to an increase of approximately $1.1$\,sec for $\lambda = 0$ and only $3.5 \times 10^{-4}$\,sec for $\lambda = -1$, respectively.
Therefore, the inverted oscillators generate the gravity-induced entanglement most efficiently and are most resistant to decoherence.

The short $\tau_{\rm ent}$ has another advantage; it opens up the possibility of generating entanglement even if $\mu$ is larger than $\eta$.
One of the main sources of the decoherence effect is interaction with air molecules~\cite{Schlosshauer2007}. To achieve $\mu_{\rm air}<\eta$, we need extremely high vacuum, $p\lesssim 10^{-17}$Pa, which is a big challenge.
However, the mean free time of the scattering with nitrogen molecules is (see 
Appendix B for detailed discussion) 
\begin{align}
    \tau_{\rm air} 
    = 0.64\, {\rm sec}\,
    \left(\frac{R}{\rm 0.2mm}\right)^{-2}\left(\frac{p}{\rm 10^{-17}Pa}\right)^{-1}
    \left(\frac{T}{\rm 1K}\right)^{\frac{1}{2}}\,,
    \label{tau air}
\end{align}
which is much longer than $\tau_{\rm ent}$ of the inverted oscillators in Eq.~\eqref{tau ent}.
Roughly requiring $\tau_{\rm air}\simeq \tau_{\rm ent}$, 
the required pressure is relaxed to $p\simeq 5.3\times 10^{-16}\,{\rm Pa}
\ \omega_{\rm kHz}$ that is within the reach of current technology~\cite{Sellner2017}.
Therefore, the inverted oscillators can potentially relax the severe requirement of decoherence suppression. 

\section{III. Example of Experimental Realization}

So far, we have examined the coupled oscillators for gravity-induced entanglement on a theoretical basis. In this section, we explore an experimental realization of inverted oscillators. Levitated optomechanical systems are suitable for achieving ultra-low decoherence by 
eliminating mechanical support, and resolving the free-fall problem mentioned in the introduction. The conventional method to prepare such a system is to levitate nanoparticles with optical tweezers~\cite{Winstone2023}. 
However, there is a limitation that the mass of the particle can be only up to nanogram scales due to the size of the trapping beam~\cite{Grier2003}. To levitate larger masses suitable for gravity experiments, optical levitation of cavity mirrors is promising~\cite{Guccione2013,Michimura2017}. It has been shown that these methods can levitate mirrors at milligram scale masses, and milligram scale is considered to be 
sweet spot for probing quantum nature of gravity~\cite{Schmole2016, Matsumoto2019,Westphal2021}. 
In such cavity optomechanical systems, inverted oscillators can be prepared by an optical anti-spring based on cavity detuning~\cite{Chen2013, Aspelmeyer2014} or on cavity geometry~\cite{Solimeno1991,Sidles2006}. We first show that optical anti-spring based on cavity detuning leads to large decoherence due to photon shot noise. We then show that geometric anti-spring can significantly suppress decoherence, and study levitated mirrors in a sandwich configuration~\cite{Michimura2017} as an example of experimental realization.

\subsection{A. Anti-spring effect of detuned cavity}

In cavity optomechanics, optical anti-spring can be generated by injecting a red-detuned laser beam.
The resonant frequency in the longitudinal direction of the anti-spring mirror due to detuning $\Delta\equiv \omega_{\ell}-\omega_{\rm cav}<0$ is~\cite{Chen2013}
\begin{equation}
    \omega_{\rm opt}^2 = 
    \frac{4\omega_{\rm \ell} P_{\rm cav}}{m c L  \kappa}
    \frac{\Delta/\kappa }{[1 + (\Delta/\kappa)^2]^2}\,,
    \label{omega opt}
\end{equation}
where $\omega_{\ell}$ is the laser frequency, $\omega_{\rm cav}$ is the cavity resonant frequency, $P_{\rm cav}$ is the intracavity power, $L$ is the cavity length, and $\kappa$ is the amplitude decay rate of the cavity. Here, we assumed that the intrinsic mechanical frequency is negligibly smaller than $|\omega_{\rm opt}|$. 

However, the quantum fluctuation of the radiation pressure of the intracavity photons, namely photon shot noise, leads to a new source of decoherence.
Its corresponding decoherence parameter is (see 
Appendix~C for derivation)
\begin{equation}
    \mu_{\rm shot}= \frac{\kappa}{|\Delta|}\,.
    \label{mu shot}
\end{equation}
Making $\mu_{\rm shot}$ smaller than $\eta$ in Eq.~\eqref{eta estimate}
would be experimentally challenging, as it would require extremely large detuning of $|\Delta|/\kappa \sim 10^{13}$. 
Even in far-detuned regimes used in current experiments, $\kappa/|\Delta|$ is typically no smaller than $10^{-3}$, which is still 10 orders of magnitude above the required level.
Moreover, from Eq.~\eqref{omega opt} such large detuning makes $\omega_{\rm opt}$ tiny, which leads to longer $\tau_{\rm ent}$. Therefore, optical anti-spring based on cavity detuning is not suitable for our purpose.

\subsection{B. Levitated mirrors in sandwich configuration}

\begin{figure}
\begin{center}
\includegraphics[width=0.9\linewidth]{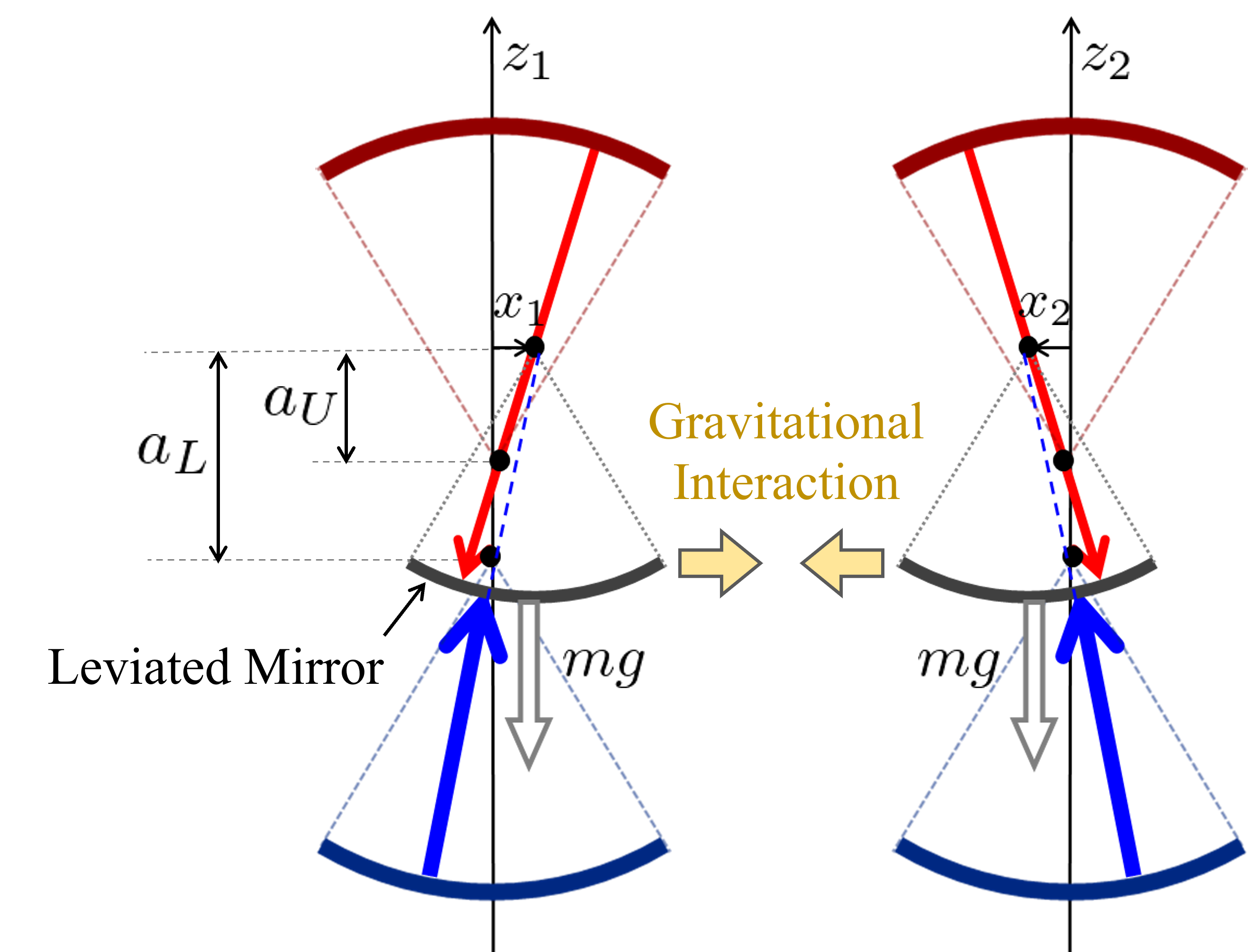}
\caption{
Our working example of an optomechanical setup. Three vertically aligned mirrors form a sandwich configuration. The upper and lower mirrors are fixed. They constitute two cavities (red and blue) that levitate the center mirror and control the stability of its horizontal motion. After preparing the initial static state of the levitated mirror in a trapping potential, one can instantly switch the optics parameters and place the levitated mirror in an unstable potential for the horizontal direction. We arrange two of these configurations side by side and let the levitated mirrors gravitationally interact until they generate detectable entanglement.}
\label{fig: Sandwich}
\end{center}
\end{figure}
To address the shot noise problem seen above, we now consider the anti-spring effect based on the cavity geometry in the transversal motion of the mirror. The shot noise can be significantly reduced, as the intracavity photons push the mirrors in a longitudinal direction.

Optical levitation of mirrors in a sandwich configuration illustrated in Fig.~\ref{fig: Sandwich} is an example of experimental realizations that can create anti-spring in the transversal motion of levitated mirrors~\cite{Michimura2017,Kawasaki2020}. 
The upper and lower mirrors are fixed and compose two Fabry-P\'erot cavities with the levitated mirror at the center.
In this setup, trapping potential for the horizontal direction can be switched instantly between stable to unstable by switching the optical parameters (see Appendix D).
To keep its levitation, we need to satisfy
\begin{equation}
    mg = \frac{2}{c}\left(P_{\rm L}-P_{\rm U}\right)\,,
    \label{vertical balance}
\end{equation}
where $g$ is the gravitational acceleration and $P_{\rm L/U}$ is 
the intracavity power of the lower/upper cavity.
Since the levitated mirror is convex downward, the lower
and upper cavity destabilizes and stabilizes its horizontal motion, respectively.
The resonant frequency of the center mirror in the horizontal direction is
\begin{align}
    &\omega^2_{\rm hor} = \frac{2}{mc}\left(
    \frac{P_{\rm U}}{a_U}-\frac{P_{\rm L}}{a_L}\right)
    =\frac{2(a_L-a_U)}{mc\, a_U a_L}P_L-\frac{g}{a_U}\,,
    \label{omehor}
    \\
    &\simeq - (1{\rm kHz})^2\left(\frac{m}{\rm 0.1mg}\right)^{-1}\left(\frac{P_L}{\rm 30kW}\right)
    \left(\frac{a_L}{\rm 2mm}\right)^{-1}\,,
    \label{omega hor}
\end{align}
where $a_{L/U}$ is the distance between the centers of curvature of the levitated mirror and the lower/upper mirror, and we used Eq.~\eqref{vertical balance} for the second equation. 
When $\omega^2_{\rm hor}$ is negative (positive), the levitated mirror behaves as an inverted (harmonic) oscillator.
To prepare the initial state in a harmonic potential and then to switch to an inverted oscillator, we can effectively change $a_U$, as described in 
Appendix~D.
The frequency of the inverted oscillator is evaluated in the second line of Eq.~\eqref{omega hor}, where we ignored the contribution from $P_U$, because the $P_L$ term dominates after the change of $a_U$.
In realizing levitated mirrors, the fabrication of $0.1$ mg curved mirrors and the suppression of photothermal effects are two major technical challenges.
The former requires the development of fabrication techniques to realize precise curvature at the milligram scale without causing the mirror to crack under the stress of high-reflectivity coatings. The latter involves mitigating thermal distortions induced by absorption of laser light, which can significantly affect the stability of the optical trap. Despite these difficulties, various experimental efforts are underway to address them~\cite{Michimura2020}.

Now we estimate the decoherence effect of the shot noise in the horizontal direction. The lateral spread of the wave function of the levitated mirror at the time of the observable entanglement generation $\tau_{\rm ent}$ is 
\begin{equation}
    \Delta x \simeq \frac{e^{\omega\tau_{\rm ent}}}{\sqrt{2m\omega_{\rm in}/\hbar}}\simeq 0.3{\rm pm}\,\omega_{\rm kHz}^{3/2}
    \left(\frac{m}{\rm 0.1mg}\right)^{-\frac{1}{2}}\left(\frac{\omega_{\rm in}}{1 \rm MHz}\right)^{-1},
    \label{Delta X}
\end{equation}
where $\omega=|\omega_{\rm hor}|$ is the frequency of the inverted potential, and $\omega_{\rm in}$ is the frequency of the initial trapping potential in which the oscillators reached the ground state. 
When this initial potential is deep $\chi\equiv \omega_{\rm in}/\omega\gg1$, the initial uncertainty~\eqref{cov mat}  reads $\sigma(0)={\rm diag}[\chi^{-1},\chi,\chi^{-1},\chi]/2$ and then $e^{\omega\tau_{\rm ent}}$ shrinks by $\sqrt{2/\chi}$ compared to Eq.~\eqref{tau ent} (see 
Appendix~A). Eq.~\eqref{Delta X} accounts for this effect.

The shot noise from the lower cavity, which is louder than the upper cavity,
is suppressed at least by a factor of $(\Delta x/a_L)^2$. 
Its decoherence effect is evaluated as the parameter $\mu$ introduced in Eq.~\eqref{mu parameter},
\begin{align}
    &\mu_{\rm shot, hor} 
    = \frac{16\omega_\ell P_L}{m\omega^2 c^2 T_{\rm in}}
    \left(\frac{\Delta x}{a_L}\right)^2
    \simeq \frac{8\omega_\ell \Delta x^2}{c a_L T_{\rm in}},\notag\\
    &= 2.5\times10^{-14}\,\omega_{\rm kHz}^{3}
    \left(\frac{a_L}{\rm 2mm}\right)^{-1}\left(\frac{m}{\rm 0.1mg}\right)^{-1}\left(\frac{\omega_{\rm in}}{\rm 1MHz}\right)^{-2},
\end{align}
where we use Eq.~\eqref{omega hor} for the second equation, the laser wavelength of $1064$nm, and the power transmittance of the cavity input mirrors $T_{\rm in}=0.1$. Compared to the single detuned cavity case~\eqref{mu shot}, the decoherence effect of the shot noise is dramatically reduced.
Note that although our setup additionally has an optical damping rate, its effect is negligibly small as discussed in 
Appendix~E.


To measure the gravity-induced entanglement, we place two copies of the sandwich configuration next to each other. We initially trap the levitated mirrors in a harmonic potential and then let the levitated mirrors gravitationally interact in the inverted potential by instantly switching the parameter $a_U$. After $\tau_{\rm ent}$ has passed, we measure their horizontal momenta and positions. This operation should be repeated for many times to significantly detect the logarithmic negativity. Our setup allows for continuous iteration of the process by returning the potential to harmonic form after the measurement, as the mirrors do not free-fall. The short $\tau_{\rm ent}$ of the inverted oscillators is also beneficial in speeding up this cycle.

\section{IV. Conclusion}

In this letter, we studied two oscillators with arbitrary quadratic potentials and calculated the gravity-induced entanglement between them. No particular gain was found by considering asymmetric oscillators. We found that the logarithmic negativity for the identical oscillators 
reduces to a simple form, $E_N \simeq 3(\eta-\mu)f_{\rm gra}$, in the long-time regime. As shown in Fig.~\ref{fig:fgra}, $f_{\rm gra}$ strongly depends on the potential form characterized by $\lambda$. Remarkably, inverted oscillators ($\lambda=-1$) have exponentially growing $f_{\rm gra}$,
which generate entanglement most efficiently and are most resistant to decoherence.
The short time scale $\tau_{\rm ent}$ of the inverted oscillators would also help to avoid decoherence from molecular collisions.

We also investigated experimental realizations of inverted oscillators in levitated optomechanics. We pointed out that the anti-spring by detuning in a single cavity suffers from quick decoherence due to photon shot noise. Then, we considered a sandwich configuration, which geometrically creates the anti-spring. It enables us to dramatically suppress shot noise decoherence because gravity acts in the horizontal direction while the laser direction is vertical.  We also noted that our optomechanical setup allows repetitive measurement cycles and solves the free-fall problem. 
Our idea of using instability to 
substantially shorten the time scale opens up new possibilities for experimental tests of the quantum nature of gravity.

\section{Acknowledgement}

We would like to thank Miles P. Blencowe for fruitful discussions.
This work was supported in part by the Japan Society for the Promotion of Science (JSPS) KAKENHI, Grants No. JP23K03424 (T.F.), JP23K13103 (A.M.), JP20H05854 (T.F.\&Y.M.), by JST PRESTO Grant No. JPMJPR200B (Y.M.), by JST CREST Grant No. JPMJCR1873 (Y.M.),
and by the Nagoya University Interdisciplinary Frontier Fellowship (Y.K.).

\section{Appendix}

\appendix
\section{A. Solution for the coupled oscillators}

With the variable $u(t)$ in Eq.~(6),
the Heisenberg-Langevin equations~(4) can be rewritten as
\begin{equation}
    \dot{u}(t)= K\, u(t)+\ell(t), \notag
\end{equation}
with
\begin{equation}
    K\equiv
    \begin{pmatrix}
    0 & \omega & 0 & 0\\
    \omega(\eta-\lambda_1) & 0 & -\omega \eta & 0\\
    0&0&0& \omega\\
    -\omega\eta &0 &\omega(\eta-\lambda_2) &0
    \end{pmatrix}\,,
    \ 
    \ell(t)=    \begin{pmatrix}
    0\\
    \xi_1\\
    0\\
    \xi_2
    \end{pmatrix}\,,\notag
\end{equation}
where we employed matrix representation.
Its analytic solution is found as
\begin{equation}
    u(t) = W_+(t)u(0)+W_+(t)\int^t_0 {\rm d}t'
    W_-(t') \ell (t')\,,\notag
\end{equation}
with $W_\pm(t) \equiv e^{\pm K t}$.
The covariance matrix~(7) is also obtained as
\begin{align}
    \sigma(t)&=W_+(t)\sigma(0)W_+^T(t)
    \notag\\
    &+W_+(t)\left[\int^t_0{\rm d}t'
    W_-(t')D W_-^T(t')\right]W_+^T(t)\,,\notag
\end{align}
with $D\equiv {\rm diag} (0,\mu\omega , 0,\mu\omega)$.
We numerically evaluate the logarithmic negativity~(10) of this solution and draw Fig.~1.

    After setting $\lambda\equiv\lambda_1=\lambda_2$, by calculating the minimum symplectic eigenvalue and expanding it with respect to small parameters $\eta\sim \mu\ll1$, we find
\begin{align}
   \tilde{\nu}_{\rm min}-\frac{1}{2} = 
    -\eta f_{\rm gra}(t)
    +\mu f_{\rm dec}(t)+\mathcal{O}(\eta^2,\mu^2)\,.
    \notag
\end{align}
The full expressions for $f_{\rm gra}$ and $f_{\rm dec}$ are given by
\begin{align}
    f_{\rm gra}&=\frac{1}{8\sqrt{2}\lambda^{3/2}}\left[\mathcal{C}_1+\mathcal{C}_2\cos (2 \sqrt{\lambda } t \omega )\right.\notag\\
    &\left.~~~~~ +\mathcal{C}_3\sin (2 \sqrt{\lambda } t \omega )+\mathcal{C}_4\cos (4 \sqrt{\lambda } t \omega )\right]^{1/2}\,,
   \notag\\
   f_{\rm dec}&=\frac{2 \sqrt{\lambda } (\lambda +1) \omega t+(\lambda -1) \sin \left(2 \sqrt{\lambda } \omega t\right)}{8 \lambda ^{3/2}}\,,
   \notag
\end{align}
with $\mathcal{C}_1=1+\lambda  \left(\lambda +8 (\lambda -1)^2  \omega^2 t^2+14\right),\,  \mathcal{C}_2=-16 \lambda,\, \mathcal{C}_3=8 \sqrt{\lambda } \left(\lambda ^2-1\right) \omega t ,$ and $ \mathcal{C}_4=-(\lambda -1)^2$.
Plugging $\lambda=0, \pm1$ and further taking the long time limit, $\omega t\gg 1$, one finds Eq.~(12).

With a different initial covariance matrix $\sigma(0)={\rm diag}[\chi^{-1},\chi,\chi^{-1},\chi]/2$,
one can repeat the same procedure and finds 
\begin{align}
      f_{\rm gra}&=\frac{1}{8 \sqrt{2} \chi}\left[
      8 t^2 \left(\chi ^2+1\right)^2 \omega ^2-\chi ^4+14 \chi^2 -1 
      \right.\notag\\
      &
      -16 \chi^2 \cosh (2 t \omega )-8\left(\chi ^4-1\right) \omega t \sinh (2  \omega t)
      \notag\\
      &\left.+\left(\chi ^2+1\right)^2 \cosh (4 \omega t)
      \right]^{1/2}\,,\notag
      \\
      f_{\rm dec}&=\frac{\left(\chi ^2+1\right) \sinh (2 \omega t)-2  \left(\chi ^2-1\right) \omega t }{8 \chi }\,,
      \notag
\end{align}
for $\lambda=-1$.
In the long time limit $\omega t\gg 1$, they read $f_{\rm gra}\simeq f_{\rm dec}\simeq (\chi^2+1)e^{2\omega t}/16\chi$ and are approximated by $\chi e^{2\omega t}/16$ for $\chi\gg 1$.
Compared to Eq.~(12), $f_{\rm gra}$ becomes larger by a factor of $\chi/2$
and thus $e^{\omega \tau_{\rm ent}}$ is smaller by $\sqrt{2/\chi}$.

\section{B. Decoherence and the Time scale of the entanglement generation}

According to the analysis of the previous section, it appeared as if the suppression of decoherence down to $\mu<\eta$ was an unavoidable condition for the entanglement generation because $E_N\simeq 3(\eta-\mu)f_{\rm gra}$ in the long time regime. 
However, this is not necessarily the case, if the entanglement generation occurs fast enough.

The main sources of the decoherence effect are typically collisions with air molecules, scattering of thermal photons and absorption (or emission) of thermal photons. 
In terms of the decoherence parameter $\mu$, these processes are expressed as
\begin{align}
    \mu_{\rm air} &= \frac{16p R^2}{3\hbar m\omega^2}\sqrt{2\pi m_{\rm air}k_B T}\,,
    \notag\\
    &=4\times 10^{-13} \omega_{\rm kHz}^{-2}
    \left(\frac{p}{\rm 10^{-17}Pa}\right)\left(\frac{T}{\rm 1K}\right)^{\frac{1}{2}}
    \left(\frac{R}{\rm 0.2mm}\right)^{2},
\notag
\\
\mu_{\rm ph} &\simeq 10^4 \frac{2\hbar c}{m\omega^2}R^6\left(\frac{k_B T}{\hbar c}\right)^9\,,
    \notag\\
    &=2\times 10^{-19} \omega_{\rm kHz}^{-2}
    \left(\frac{T}{\rm 1K}\right)^{9}
    \left(\frac{R}{\rm 0.2mm}\right)^{6},
\notag
\\
\mu_{\rm abs} &\simeq 10^3 \frac{\hbar c}{2m\omega^2}R^3\left(\frac{k_B T}{\hbar c}\right)^6\,,
    \notag\\
    &=9\times 10^{-19} \omega_{\rm kHz}^{-2}
    \left(\frac{T}{\rm 1K}\right)^{6}
    \left(\frac{R}{\rm 0.2mm}\right)^{3},
\notag
\end{align}
where $R$ is the radius of the oscillator, $p$ and $T$ are the pressure and temperature of the environment, and we used the mass of nitrogen molecule $m_{\rm air}=4.7\times 10^{-23}$g and $m=0.1$mg.
For simplicity, we approximated the dielectric constant factor by unity. For our fiducial parameters used above, the air molecule scattering is the leading decoherence process, $\mu_{\rm air}\gg \mu_{\rm ph}, \mu_{\rm abs}$.
To achieve $\mu_{\rm air}<\eta$, we need extremely high vacuum, $p\lesssim 10^{-17}$Pa.
Thus, the sufficient suppression of decoherence 
is a major challenge.

Nonetheless, even if $\mu_{\rm air}$ is larger than $\eta$, decoherence should be ineffective, if no scattering occurs during the experiment. How often do the air molecules actually hit the oscillator?
The mean free time of the scattering between the oscillator and the molecules 
is evaluated as
\begin{align}
    \tau_{\rm air} &= (\pi R^2 v_x n_{\rm air})^{-1}\,, \notag\\
    &= 0.64\, {\rm sec}\,
    \left(\frac{R}{\rm 0.2mm}\right)^{-2}\left(\frac{p}{\rm 10^{-17}Pa}\right)^{-1}
    \left(\frac{T}{\rm 1K}\right)^{\frac{1}{2}}\,,
    \notag
\end{align}
where $n_{\rm air}=p/k_B T$ is the molecule density and $v_x=\sqrt{k_B T/m_{\rm air}}$ is its velocity in one direction.
Note that $\tau_{\rm air}$ is much longer than $\tau_{\rm ent}$ of the inverted oscillators in Eq.~(13).
Roughly requiring $\tau_{\rm air}\gtrsim \tau_{\rm ent}$, we find the maximum allowed value of the pressure for the inverted oscillators as
\begin{equation}
p\lesssim 5.3\times 10^{-16}\,{\rm Pa}
\ \omega_{\rm kHz}\left(\frac{R}{\rm 0.2mm}\right)^{-2}\left(\frac{T}{\rm 1K}\right)^{\frac{1}{2}}.
\notag
\end{equation}
Extremely low pressure at the level of $p\simeq 10^{-16}$Pa has been experimentally achieved.
Although the higher temperature is apparently favorable, other decoherence processes such as thermal photon scattering would be significant.
Consequently, the inverted oscillators can potentially relax the ultrahigh vacuum requirement and make the experiment more realistic by increasing $\omega$.
For a more rigorous treatment, one should revisit the Heisenberg-Langevin equations~(4) and go beyond the Markovian random force approximation of $\xi_i$.

\section{C. Decoherence from shot noise in detuned cavities}

The intracavity power fluctuates due to the intrinsic shot noise of the laser beam as
\begin{equation}
    \delta P_{\rm cav}=\sqrt{2\hbar \omega_\ell P_{\rm in}}\frac{P_{\rm cav}}{P_{\rm in}},
    \notag
\end{equation}
where $P_{\rm cav}= 4P_{\rm in}/(T_{\rm in}[1+(\Delta/\kappa)^2])$ is the intracavity power. The corresponding one-sided power spectrum density of the force is
\begin{equation}
    S_{\rm shot}^F=\left(\frac{2\delta P_{\rm cav}}{c}\right)^2
    =\frac{32\hbar \omega_{\ell} P_{\rm cav}}{c^2 T_{\rm in}[1+(\Delta/\kappa)^2]}\,.
    \notag
\end{equation}
The decoherence parameter $\mu$ can be computed as
\begin{equation}
    \mu_{\rm shot} = \frac{S^F_{\rm shot}}{2\hbar m\omega^2}
    =\frac{16 \omega_{\ell} P_{\rm cav}}
    {\omega^2 m  c^2T_{\rm in}[1+(\Delta/\kappa)^2]} .
    \notag
\end{equation}
Substituting $|\omega_{\rm opt}|$ of Eq.~(15) into $\omega$ in the above equation and using $T_{\rm in}=4L\kappa/c$, which ignores the intracavity loss, we obtain Eq.~(16).
Since both $\omega_{\rm opt}^2$ and $S_{\rm shot}^F$ scales linearly with optomechanical coupling strength, Eq.~(16) will be only dependent on normalized detuning. 

\section{D. Setup for switching potentials}

The expression for the frequency of the levitated mirror in the horizontal direction $\omega_{\rm hor}^2$ in Eq.~(18) implies that changing the intracavity power $P_L$ alone cannot convert a harmonic potential to a high-frequency inverted potential, because the coefficient of $P_L$ remains negative for $a_L<a_U$.
To prepare the initial state and start the evolution of the coupled oscillators, it is crucial that our system can provide both harmonic and inverted potentials, as well as quickly switch between them. 

\begin{figure}
\begin{center}
\includegraphics[height=8.0cm]{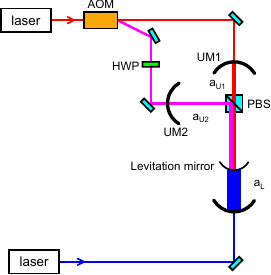}
\caption{Example setup for effectively changing $a_U$. AOM: acousto-optic modulator, HWP: half-wave plate, PBS: polarizing beamsplitter.}
\label{fig: initialstate}
\end{center}
\end{figure}

This can be done, for example, by an experimental setup illustrated in Fig.~\ref{fig: initialstate}. There are two upper cavities formed by an upper mirror labeled UM1(UM2) and the levitation mirror. The first upper cavity has a distance between the centers of curvature of $a_{U1}$ which meets $a_L > a_{U1}$, to create a harmonic potential, and the second upper cavity has $a_{U2}$ which meets $a_L < a_{U2}$, to create an inverted potential. Two cavities have different polarizations to avoid coupling between two cavities, and a polarizing beamsplitter is inserted in the cavity to share the same levitation mirror. By switching relative intracavity power between the first and the second upper cavities by an acousto-optic modulator (AOM), we can effectively switch between harmonic and inverted potentials. The typical speed for the switch is less than 100~nsec for an AOM, which is much faster than the time required to generate the entanglement in Eq.~(13).
%

\section{E. Optical damping rate of the sandwich configuration}

In our sandwich coordination, the optical effect not only gives the spring constant of Eq.~(18), but also introduces an additional optical damping rate.
This is because the change in the intracavity power due to the mirror motion is not instantaneous.
Here, we show that the damping effect is negligibly small.

The expression for the optical damping rate in the horizontal direction is given by~[31]
\begin{equation}
    \gamma_I^{\rm hor} = \frac{T_{\rm in} P_{I}}{mc^2} \frac{\ell_I/a_I}{1-G_I}\,,
    \quad (I=U,L)
    \notag
\end{equation}
where $\ell_I$ is the cavity length, $G_I=(1-\ell_I/R_I)(1-\ell_I/R_C)$, and $R_I$ is the radius of curvature of the
upper, center, lower mirror for $I=U,C,L$, respectively.
Although the total damping ratio is the sum of the two $\gamma_{\rm hor}=\gamma^{\rm hor}_U+\gamma^{\rm hor}_L$, 
we focus on $\gamma^{\rm hor}_L$ because they are proportional to $P_I/a_I$ and the lower contribution is dominant as in Eq.~(19).
Assuming the cavity length and the radius of curvature are the same order and ignoring the second factor, the above equation reads
\begin{equation}
    \gamma_{\rm hor} \simeq \frac{T_{\rm in} P_L}{mc^2}
    = 3\times 10^{-7}{\rm Hz}\,
    \nonumber 
\end{equation}
where we used the same parameters as Eq.~(19). Since this value is much smaller than the resonant frequency $|\omega_{\rm hor}|=1$kHz, we can safely ignore this optical damping effect.

\bibliographystyle{unsrt}
\bibliography{reference}

\end{document}